\documentclass[12pt,reprint,aps,groupedaddress,nofootinbib,prd,twocolumn]{revtex4-2}
\usepackage[utf8]{inputenc}
\usepackage{amsmath,amssymb,graphicx,amsfonts}
\usepackage{mathrsfs}
\usepackage{comment}
\usepackage{xcolor}
\usepackage[shortlabels]{enumitem}
\usepackage{hyperref}
\usepackage{url}

\begin{document}
\title{Charged black holes in de Sitter space: superradiant amplification of charged scalar waves and resonant hyperradiation}

\author{Giacomo Mascher$^{1,2}$, Kyriakos Destounis$^1$ and Kostas D. Kokkotas$^1$} 
\affiliation{$^1$Theoretical Astrophysics, IAAT, University of T{\"u}bingen, 72076 T{\"u}bingen, Germany}
\affiliation{$^2$Physics Department, University of Trento, Via Sommarive 14, 38123 Trento, Italy}

\begin{abstract}
	Black holes typically inhabit highly dynamical galactic environments and are frequently permeated by accretion media. The inevitable scattering of scalar, electromagnetic and gravitational waves off rotating and charged black holes provides a remarkable demonstration of the energy extraction and subsequent amplification of scattered waves under the expense of the compact object's rotational and electromagnetic energy; a phenomenon known as superradiance. Certain circumstances tremendously favor the energy extraction process in such extent that linear superradiant instabilities are triggered. We examine the superradiant amplification of monochromatic charged scalar fields impinging Reissner-Nordstr\"om-de Sitter black holes which are known to exhibit superradiantly unstable quasinormal modes. We find that even though the frequency range of superradiance is reduced when a positive cosmological constant is incorporated, the amplification factors are significantly elevated with respect to those occurring in Reissner-Nordstr\"om spacetime. Intriguingly, we confirm that the long-lived quasinormal resonances that reside in the superradiant regime induce resonant peaks when the wave's frequency matches the real part of the quasinormal mode, regardless of the spacetime's modal stability. 
\end{abstract}
\maketitle

\section{Introduction}

Gravitational-wave (GW) astrophysics is in full bloom since the first direct detection of gravitational radiation from the infamous GW150914 black hole (BH) merger \cite{LIGOScientific:2016aoc}. The Laser Interferometer Gravitational-Wave Observatory (LIGO) and Virgo continue detecting GW events with imperious momentum \cite{LIGOScientific:2021djp}, which has lead to a vast database of incoming merger signals from the edges of our Cosmos. 

Our current knowledge regarding the evolution of gravitationally captured compact objects indicates that the merger's constituents undergo a decaying circumrotation around a center of mass, know as the inspiral stage, till they merge in a violent manner to unite into a single remnant which vibrates till its relaxation to a final stable state is achieved. 

The final phase is known as the ringdown and can be, in principle, characterized by the vibrational spectra of the final perturbed compact object; the quasinormal mode (QNMs) frequencies \cite{Kokkotas:1999bd,Berti:2009kk,Konoplya:2011qq}. The extraction of QNMs from realistic GW ringdown data, in order to characterize the externally observable properties of the final object, the underlying spacetime geometry and test the no-hair theorem of General Relativity (GR) is currently one of the most thriving topics of data-analysis exploration, known as BH spectroscopy \cite{Echeverria:1989hg,Dreyer:2003bv,Berti:2005ys,Berti:2007zu,Baibhav:2017jhs,Giesler:2019uxc,Isi:2019aib,Bustillo:2020buq,Isi:2021iql,Cotesta:2022pci}. 

Besides the fact that BHs appear as `gravitational' bells, they exhibit a myriad of interesting phenomena which have yet to be detected experimentally. BHs are not completely dark and emit Hawking radiation \cite{Hawking:1974rv,Hawking:1975vcx}. Most importantly, BHs are sources of energy. They can amplify incident waves that scatter off them, under certain circumstances, in the expense of their rotational or electromagnetic energy \cite{Penrose:1971uk,Bekenstein:1973mi}. This phenomenon is similar to the Penrose process \cite{Penrose:1971uk} and is known as superradiance \cite{Zeldovich}. The vast literature of BH superradiance covers rotating and charged BHs, as well as stars and compact objects in GR and various modified theories of gravity (see the review \cite{Brito:2015oca} for further details).

The epitome of superradiance in GR are Reissner-Nordstr\"om (RN), Kerr and Kerr-Newman BHs \cite{DiMenza:2014vpa,Benone:2014qaa,Benone:2015bst,Balakumar:2020gli,Glampedakis:2001cx,Dolan:2008kf,Benone:2019all}. In principle, the occurence of superradiance does not necessarily imply that the spacetime itself will be destabilized. Nonetheless, BH superradiant instabilities do appear when these waves are confined in negative potential wells \cite{Dolan:2007mj,Dolan:2012yt,Vitor5,RN1,Kerr3,KN1,Konoplya:2008au,Konoplya:2013sba,Vieira:2021nha,Franzin:2021kvj} or when the boundary conditions are altered \cite{Vitor1,Vitor3,Vitor4,RN5,Kerr1,Kerr2,Kerr4,Kerr5,Li:2012rx,Li:2014fna,Li:2014gfg,Li:2015mqa,Herdeiro:2013pia,Herdeiro:2018wub,Sanchis-Gual:2015lje,Sanchis-Gual:2016tcm,Cuadros-Melgar:2021sjy}. Under such circumstances, perturbations grow exponentially and lead to the formation of BH bombs \cite{Press:1972zz}.

Interestingly, RN BHs embedded in de Sitter space can superradiantly amplify perturbations and when particular conditions are met the corresponding charged scalar QNMs become unstable \cite{Zhu:2014sya,Konoplya:2014lha,Cardoso:2018nvb,Dias:2018ufh,Mo:2018nnu,Destounis:2019hca} inciting charged-de Sitter BH bombs. This superradiant instability appears in a small parameter space of the system when the cosmological constant of spacetime and the charge of spherically-symmetric scalar perturbations are sufficiently small \cite{Dias:2018ufh,Destounis:2019hca}. de Sitter space provides a physical representation of the current cosmological evolution of the Universe, with the positive cosmological constant, although minuscule, supplying ample driving force for the accelerated expansion of our Universe.

The superradiant amplification of incident waves scattering off of Reissner-Nordstr\"om-de Sitter (RNdS) BHs has not been consistently studied yet, in contrast to superradiance in Kerr-de Sitter BHs \cite{Maeda:1993}. The analysis in \cite{Maeda:1993} has shown that despite the fact that the superradiant frequency range is shortened when a positive cosmological constant is incorporated, the amplification factors of incident waves are significantly elevated with respect to those occurring in Kerr BHs. 

Our work aims to analyze the phenomenon of superradiance of charged massless and massive scalar incident waves impinging RNdS BHs with numerical scattering techniques, compare the amplification factors with those of RN BHs and interpret the effect of QNM resonances which belong in the superradiant regime. 

We find that the existence of superradiant monopolar QNMs induces strong amplification of incident waves, when the wave's frequency coincides with the real part of the QNM, and is loosely related to the imaginary part of the QNM \cite{ChandraFerrari1991,ChandraFerrari2,Kokkotas1994}. The `hyperradiant' peaks formed around QNMs are oblivious to the linear stability properties of spacetime and are consistent with Breit-Wigner resonances \cite{Thorne1969,Berti:2009wx}. When monopolar scalar waves are highly charged their amplification factors approach $100\%$ from above, since the lifetime of QNMs becomes shorter, in direct contrast to scattered waves in RN which tend to $100\%$ amplification from below \cite{Brito:2015oca}. 

We further find that higher multipole incident waves still possess superradiant frequency regimes but no QNMs, and subsequent resonant peaks, appear in such range. Even so, these waves can reach amplification factors up to $100\%$ from below at the large scalar charge limit. \emph{In finality, we demonstrate that the addition of mass to the scalar wave acts as a friction medium which rapidly diminishes resonant peaks and minimizes the amplification factors, since the increment of mass expels QNMs from the frequency domain of superradiance} \cite{Cardoso:2018nvb,Destounis:2019hca}. In what follows, we assume geometrized units such that $G=c=1$.

\section{The Reissner-Nordstr\"om-de Sitter spacetime}

RNdS BHs are spherically-symmetric static solutions to the Einstein-Maxwell field equations with a positive cosmological constant \cite{Hawking:1973uf}. They are described by the line element 
\begin{equation}
	ds^2=-f(r)dt^2+f(r)^{-1}dr^2+r^2\left(d\theta^2+\sin^2\theta\,d\phi^2\right),
\end{equation}
where
\begin{align}
	f(r)&=1-\frac{2M}{r}+\frac{Q^2}{r^2}-\frac{\Lambda r^2}{3},
\end{align}
with $M,\, Q$ the mass and charge of the BH and $\Lambda>0$ the cosmological constant. The associated potential sourced by the electric field is $A_\mu=\left(-Q/r,0,0,0\right)$.

The causal structure of a subextremal RNdS spacetime possesses three distinct null hypersurfaces, namely the Cauchy $r=r_-$, event $r=r_+$ and cosmological horizon $r=r_c$, where $r_-<r_+<r_c$. Obviously, the BH's interior lies in $r\leq r_+$, with the Cauchy horizon being the boundary of maximal globally-hyperbolic development of initial data on a Cauchy hypersurface \cite{Hawking:1973uf}. The static (observable) region lies between the event and cosmological horizons, such as $r_+<r<r_c$. In the cosmological region, where $r\geq r_c$, the accelerated expansion of the Universe is so rapid that all events are infinitely redshifted and causally disconnected with the static region.

\section{Charged massive scalar wave equation}

We are interested in the propagation of linear massive and charged scalar fields $\psi$ on a fixed RNdS background. Their motion is governed by the Klein-Gordon equation
\begin{equation}\label{KG}
	(D^\nu D_\nu-\mu^2)\psi=0, 
\end{equation}
where $\mu$, $q$ the mass and charge of the field, respectively, and $D_\nu=\nabla_\nu-iqA_\nu$ the covariant derivative which incorporates the effects of curvature and electromagnetism. By expanding the scalar field $\psi(t,r,\theta,\phi)$ in terms of spherical harmonics (due to spherical symmetry) with a harmonic time dependence (due to a Fourier decomposition) as
\begin{equation}
\psi(t,r,\theta,\phi)=\sum_{l,m}\frac{\Psi_{l m}(r)}{r}Y_{lm}(\theta,\phi)e^{-i\omega t},
\end{equation}
we reduce Eq. \eqref{KG} to a one-dimensional Schr\"odinger-like equation
\begin{equation}
	\label{master_eq_RNdS}
	\frac{d^2 \Psi}{d r_*^2}+\left[\omega^2-2\omega\Phi(r)-V(r)\right]\Psi=0\,,
\end{equation}
where $\Phi(r)=q Q/r$ is the electrostatic potential energy, $qQ$ is the charge coupling and $dr_*=dr/f(r)$ is the tortoise coordinate. The effective potential for massive and charged scalar perturbations is then
\begin{equation}
	\label{RNdS_general potential}
	V(r)=f(r)\left(\mu^2+\frac{\ell(\ell+1)}{r^2}+\frac{f^\prime(r)}{r}\right)-\Phi(r)^2,
\end{equation}
where $\ell$ is the multipolar degree (angular number) of spherical harmonics which correspond to the eigenvalue $\ell(\ell+1)$ of the squared orbital angular momentum operator. Here, primes denote derivatives with respect to the radial coordinate $r$. 

\begin{figure*}[t]
	\includegraphics[scale=0.5]{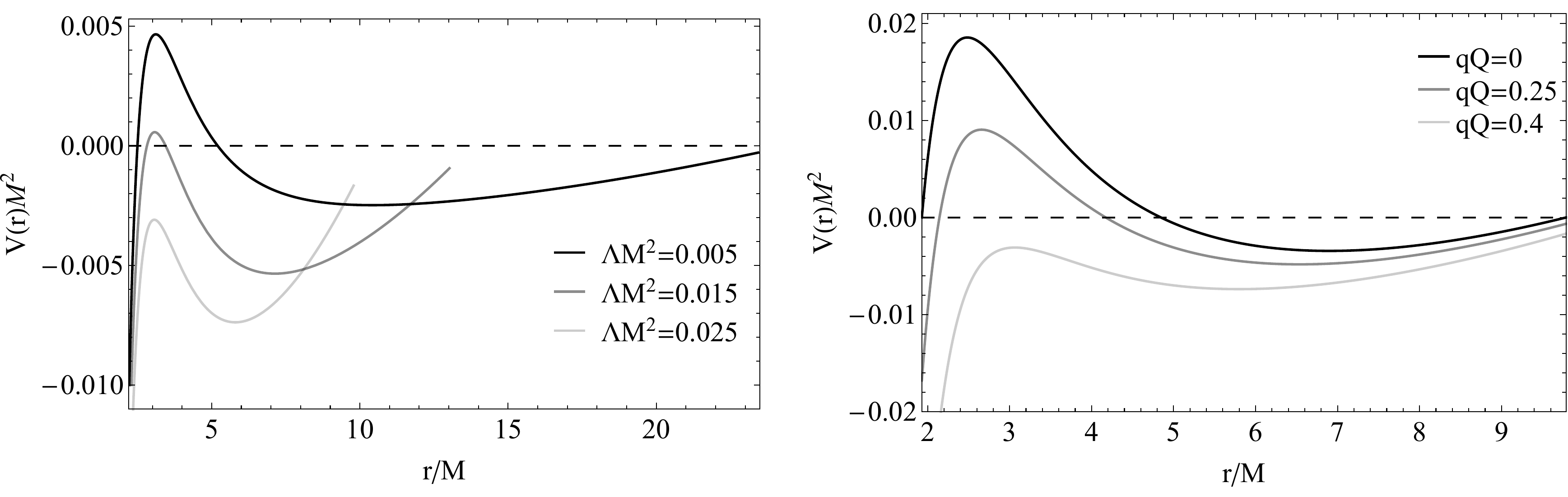}
	\caption{Left: Effective potential of $\ell=0$ massless charged scalar perturbations with charge coupling $qQ=0.4$ for a RNdS BH with charge  $Q=0.5Q_\text{max}$ and varying cosmological constants $\Lambda M^2$. Right: Effective potential of $\ell=0$ massless charged scalar perturbations for a RNdS BH with charge  $Q=0.5Q_\text{max}$, cosmological constant $\Lambda M^2=0.025$ and varying charge couplings $qQ$.}
	\label{potentials}
\end{figure*}

\section{Superradiant amplification of monochromatic waves}

Let a massive and charged monochromatic incident wave from the cosmological horizon, with amplitude coefficient $\mathcal{I}$, that scatters off of the RNdS photon sphere. The wave will be partially reflected back toward the cosmological horizon, with reflection coefficient $\mathcal{R}$, and partially transmitted through the potential barrier and into the event horizon, with transmission coefficient $\mathcal{T}$. The above scattering experiment translates to the following boundary conditions in (\ref{master_eq_RNdS}):
\begin{equation}
	\label{scat}
	\Psi \sim
	\left\{
	\begin{array}{lcl}
		\mathcal{T} e^{-i (\omega-\Phi(r_+))r_* },\,\,\,\,\,\,\quad\quad\quad\quad\quad\quad r \rightarrow r_+, \\
		&
		&
		\\
		\mathcal{I}e^{-i(\omega-\Phi(r_c))r_*} + \mathcal{R} e^{i(\omega-\Phi(r_c))r_*},\, r \rightarrow r_c.
	\end{array}
	\right.
\end{equation}

Due to the fact that the Wronskian of $\Psi$ and its linearly-independent complex conjugate counterpart $\Psi^\dagger$ is $r_*$-independent, we immediately realize that the Wronskian at both boundaries coincide. This equality leads to the following relation
\begin{equation}\label{Wronskian}
	|\mathcal{R}|^2=|\mathcal{I}|^2-\frac{\omega-\Phi(r_+)}{\omega-\Phi(r_c)}|\mathcal{T}|^2.
\end{equation}
From relation \eqref{Wronskian}, we observe that when the incident wave's frequency satisfies
\begin{equation}
	\label{suprad}
	\Phi(r_c)<\omega<\Phi(r_+),
\end{equation}
then the amplitude of the reflected wave is larger than the amplitude of the incident wave. Equation \eqref{suprad} is known as the superradiance relation. Physically, this translates to the amplification of incident massive and charged monochromatic waves under the expense of the BH's electromagnetic energy. Notice that the presence of a non-zero mass to the incident wave does not affect Eq. \eqref{suprad} in contrast to the case when $\Lambda M^2\rightarrow 0$, where Eq. (\ref{suprad}) reduces to the respective one for asymptotically flat RN BHs and the scalar mass becomes the lower frequency bound for superradiant amplification \cite{Bekenstein:1973mi,Brito:2015oca}.

In what follows, we use the energy fluxes of scalar fields at the cosmological horizon to define the amplification factor \cite{Brito:2015oca}
\begin{equation}\label{amplification_factor}
	Z_\ell=\frac{|\mathcal{R}|^2}{|\mathcal{I}|^2}-1,
\end{equation}
as a function of the frequency $\omega$. It is obvious that when $Z_\ell< 0$ superradiance does not occur while when $Z_\ell> 0$ the incident wave is superradiantly amplified. In turn, $Z_\ell=0$ holds at the bounds of the superradiant relation \eqref{suprad}.

To numerically integrate Eq. \eqref{master_eq_RNdS} with boundary conditions \eqref{scat}, we expand its solutions at the event and cosmological horizons to arbitrary order (till we reach numerical convergence) and then match the two asymptotic solutions at an intermediate point by imposing regularity of the solutions and their derivatives. Through this numerical process we extract the amplification factor for numerous monochromatic incident waves with varying $\omega$. We have performed immense convergence tests by increasing the order of expansion of solutions at the boundaries of integration and also find very good agreement with the amplification factors of RN shown in \cite{Brito:2015oca}.

\section{Quasinormal modes}

\begin{figure*}[t]
	\includegraphics[scale=0.43]{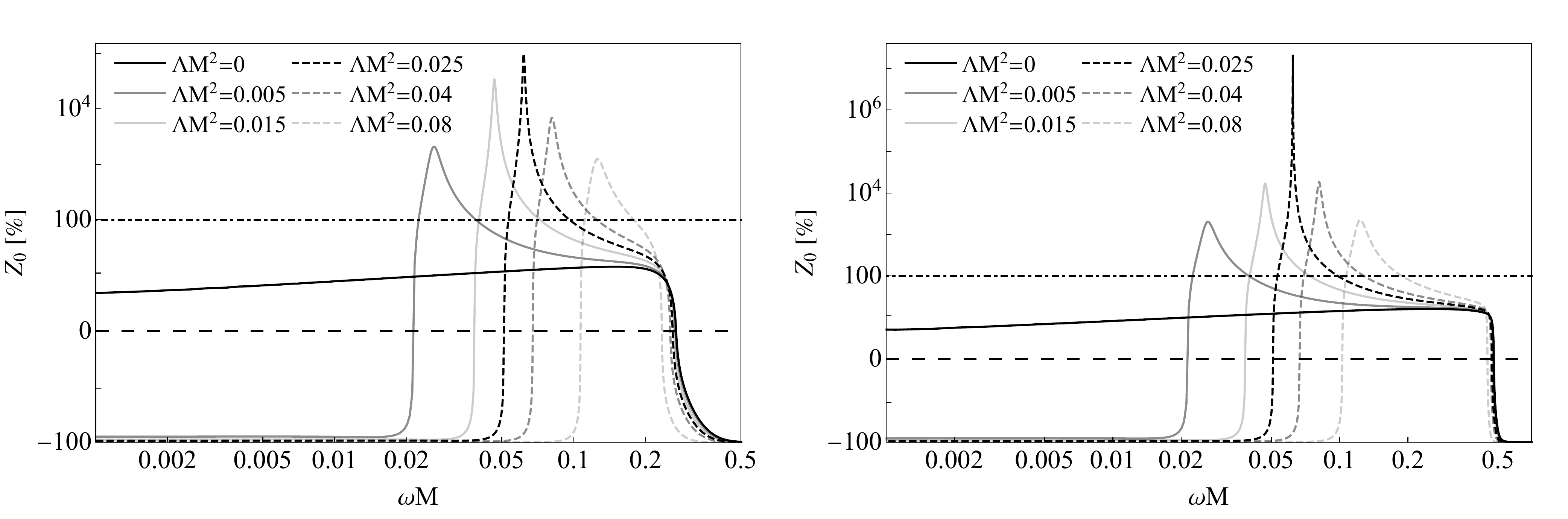}
	\caption{Left: Amplification factors of massless $\ell=0$ monochromatic waves with $qQ=0.5$ for a RNdS BH with $Q=0.5Q_\text{max}$ and varying cosmological constant $\Lambda M^2$. The horizontal black dashed line designates the onset of superradiant amplification while the black dot-dashed line designates amplification factors that equal $100\%$. Right: Same as left but with $Q=0.999Q_\text{max}$.}
	\label{varL}
\end{figure*}

\begin{figure*}[t]
	\includegraphics[scale=0.43]{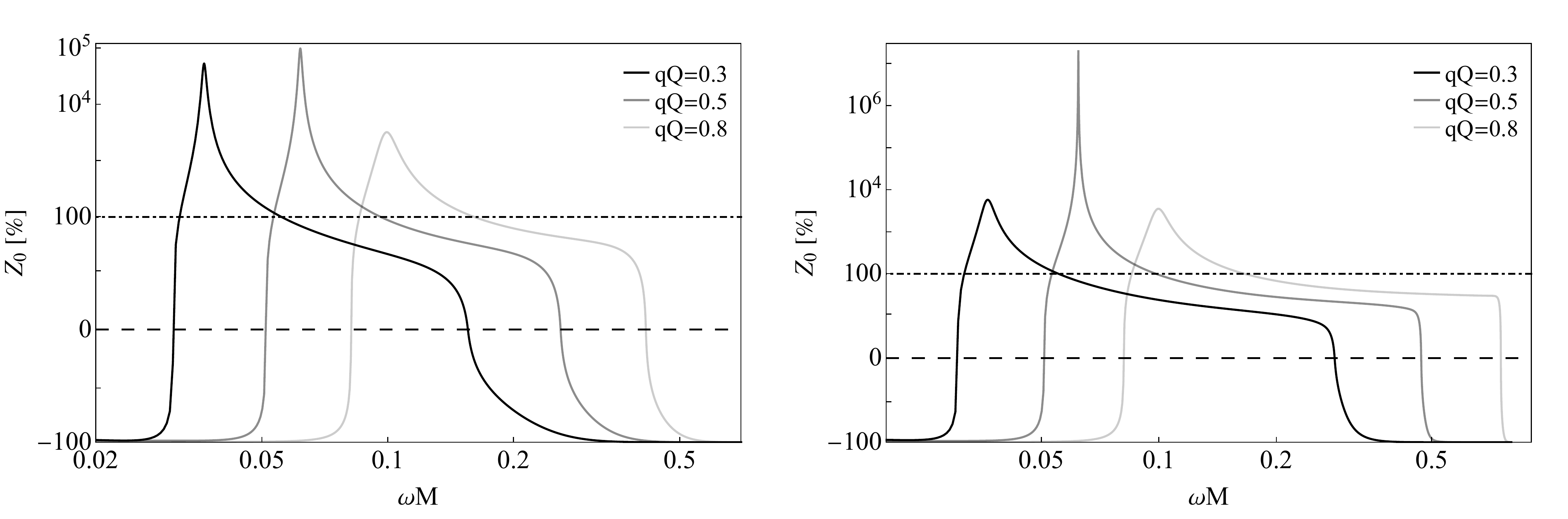}
	\caption{Left: Amplification factors of massless $\ell=0$ monochromatic waves for a RNdS BH with $Q=0.5Q_\text{max}$, $\Lambda M^2=0.025$ and varying charge coupling $qQ$. The horizontal black dashed line designates the onset of superradiant amplification while the black dot-dashed line designates amplification factors that equal $100\%$. Right: Same as left but with $Q=0.999Q_\text{max}$.}
	\label{varqQ}
\end{figure*}

To obtain the characteristic QNM frequencies $\omega_\text{QNM}$ of RNdS spacetime, we impose ingoing waves at the event horizon and outgoing waves at the cosmological horizon by neglecting the incident wave in Eq. \eqref{scat}, that is $|\mathcal{I}|=0$ for QNMs. From the asymptotic behavior of Eq. \eqref{master_eq_RNdS} these boundary conditions translate to~\cite{Berti:2009kk}
\begin{equation}
	\label{bcs}
	\Psi \sim
	\left\{
	\begin{array}{lcl}
		e^{-i (\omega-\Phi(r_+))r_* },\,\,\,\quad r \rightarrow r_+, \\
		&
		&
		\\
		e^{+i(\omega-\Phi(r_c))r_*},\,\,\,\,\quad r \rightarrow r_c.
	\end{array}
	\right.
\end{equation}
The QNM frequencies are characterized, for each $\ell$, by an integer $n\geq 0$ labeling the mode number, with the fundamental mode $n=0$ corresponding, by definition, to the non-vanishing frequency $\omega_R$ with the smallest (by absolute value) imaginary part $\omega_I$ and $n>0$ denotes higher harmonics of oscillation (overtones). Due to the underlying symmetry $\omega\rightarrow-\omega$ and $\Phi(r)\rightarrow-\Phi(r)$ of Eq. (\ref{master_eq_RNdS}), we will consider only cases where the charge coupling satisfies $qQ>0$.

The QNMs of RNdS spacetime has been extensively analyzed in the current literature. Their spectra consist of three distinct families of modes which govern the early and late-time dynamics of perturbation evolution (see \cite{Cardoso:2017soq,Cardoso:2018nvb,Destounis:2018qnb,Liu:2019lon,Destounis:2019omd,Dias:2018etb,Dias:2018ufh,Mo:2018nnu,Joykutty:2021fgj} for further details). Even though a QNM analysis is not the main point of this work, any QNMs discussed henceforth will be obtained via the Mathematica package of \cite{Jansen:2017oag} (based on collocation methods developed in \cite{Dias:2010eu}), and checked in various cases with a WKB approximation \cite{Iyer:1986np} and a code developed based on the matrix method in \cite{KaiLin1}. 

\subsection{Superradiant instability and the role of the effective potential}

The fact that incident waves can be superradiantly amplified when scattered off of charged BHs is closely connected with the properties of the underlying effective potential. In RNdS spacetime, spherically-symmetric ($\ell=0$) charge scalar perturbations form quasibound states \cite{Dolan:2007mj,Vieira:2016ubt,Vieira:2021doo,Vieira:2021xqw,Vieira:2021ozg} localized at a potential well which is present right outside the photon sphere (see Fig. \ref{potentials}). For a particular subspace of the parameter space of the system, it has been shown that such modes become unstable due to confinement and simultaneously satisfy the superradiant relation \eqref{suprad} \cite{Zhu:2014sya,Konoplya:2014lha,Cardoso:2018nvb,Destounis:2019hca}. This does not occur in RN geometries, since there are no unstable or superradiant QNMs. Thus RNdS BHs withstand an alternative charged BH bomb mechanism, under charged scalar perturbations, without the need of AdS asymptotics or artificial mirrors placed around charged BHs to confine perturbations and trigger instabilities.

\section{Amplification factors, superradiance and resonances}

In what follows we present the amplification factors $Z_\ell$ of incident charged massless and massive scalar waves scattering off of RNdS BHs.

\subsection{$\ell=0$ scalar waves} 

\begin{figure}[t]
	\includegraphics[scale=0.42]{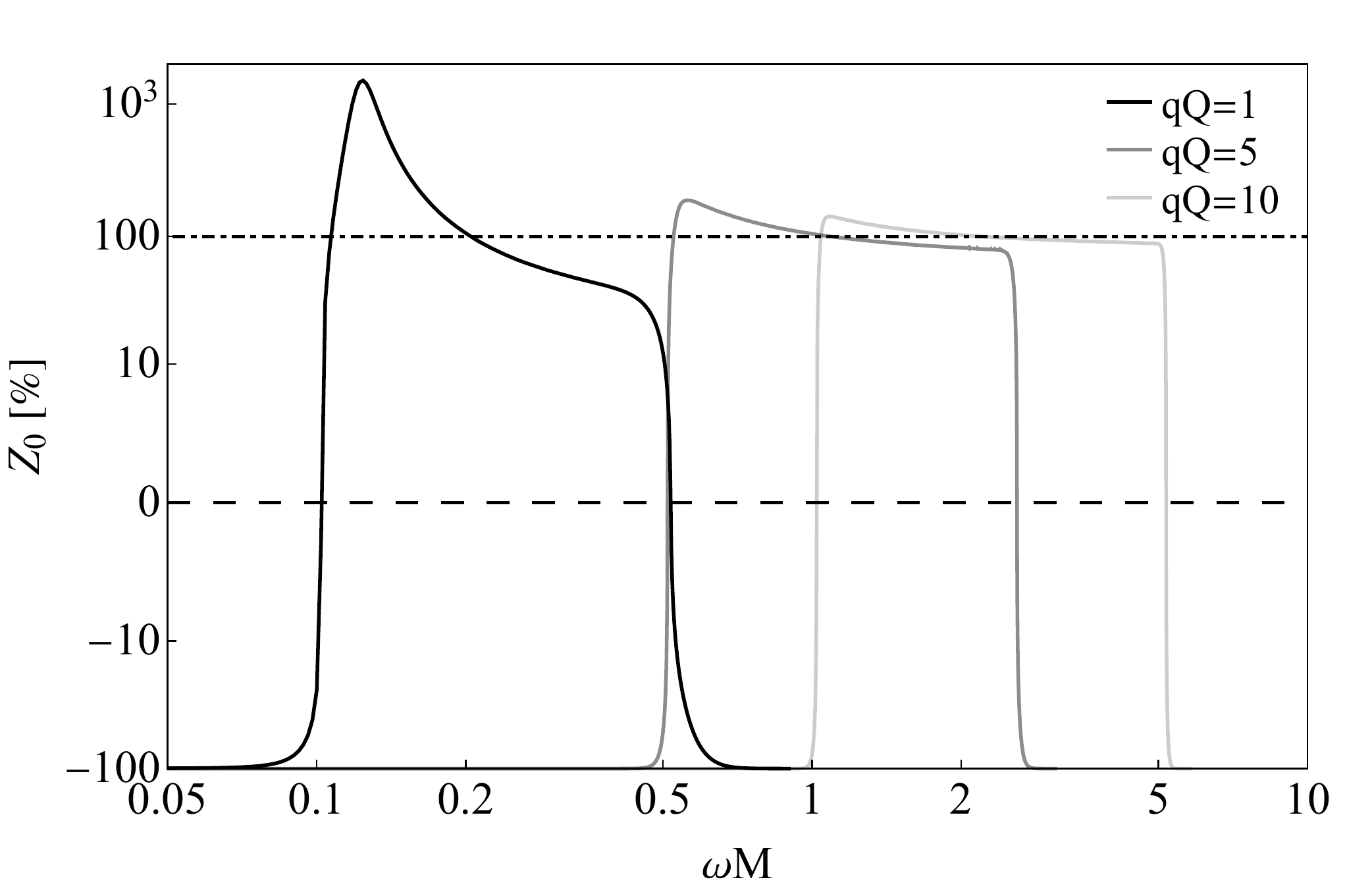}
	\caption{Amplification factors of massless $\ell=0$ monochromatic waves for a RNdS BH with $Q=0.5Q_\text{max}$, $\Lambda M^2=0.025$ and large charge couplings $qQ$. The horizontal black dashed line designates the onset of superradiant amplification while the black dot-dashed line designates amplification factors that equal $100\%$.}
	\label{largeqQ}
\end{figure}

\begin{figure}[t]
	\includegraphics[scale=0.42]{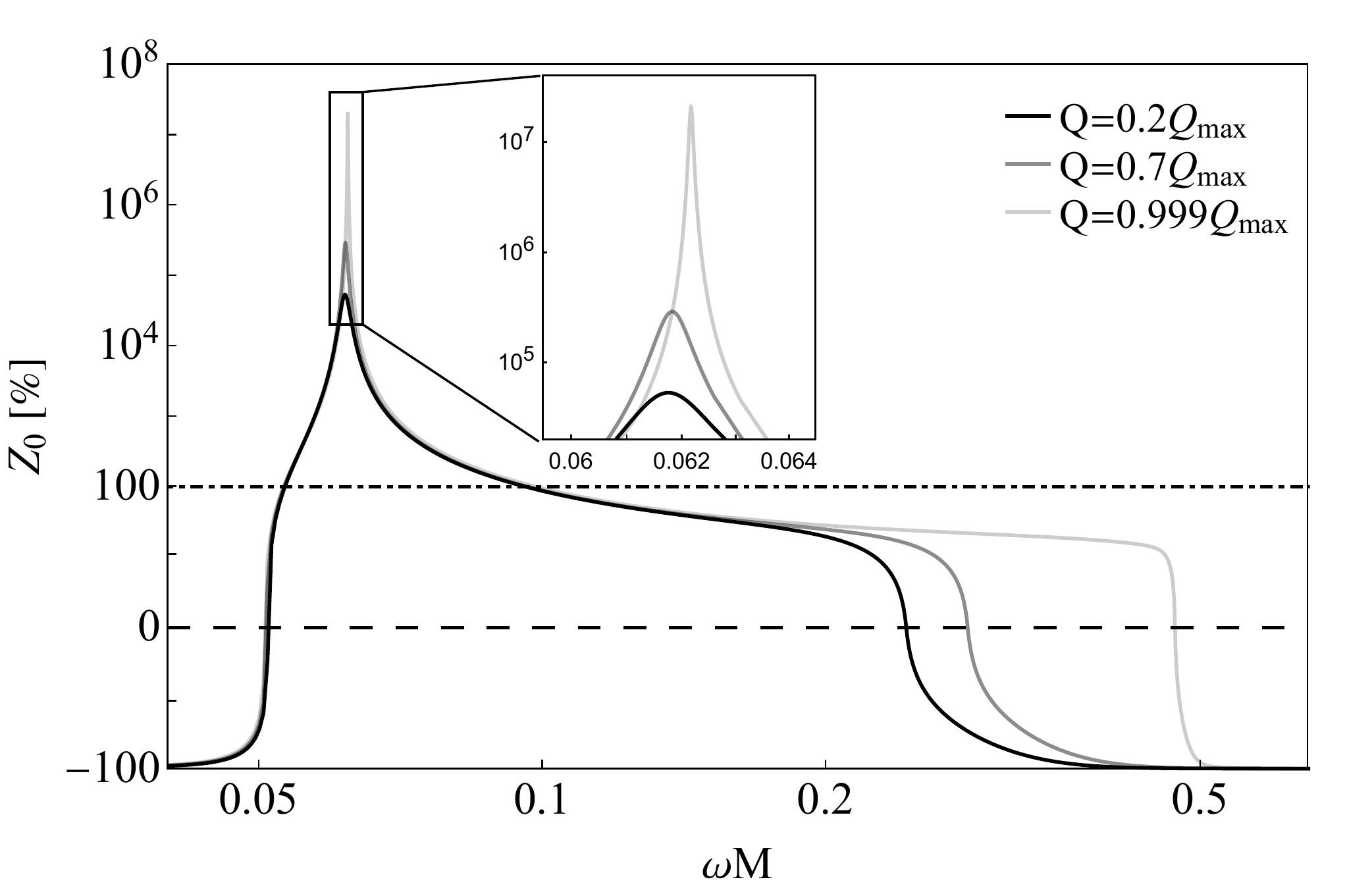}
	\caption{Amplification factors of massless $\ell=0$ monochromatic waves with charge coupling $qQ=0.5$ for a RNdS BH with $\Lambda M^2=0.025$ and varying charge $Q/Q_\text{max}$. The horizontal black dashed line designates the onset of superradiant amplification while the black dot-dashed line designates amplification factors that equal $100\%$.}
	\label{varQ}
\end{figure}

\begin{table}[t]
	\centering
	\scalebox{1.1}{
		\begin{tabular}{||c| c | c ||} 
			\hline
			\multicolumn{3}{||c||}{$Q=0.5Q_\text{max}$, $qQ=0.5$} \\
			\hline
			$\Lambda M^2$ & $\omega_\text{QNM}$ & $\omega_\text{peak}$  \\ [0.5ex] 
			\hline
			$0.015$ & 0.04652 + 0.00074 i & 0.04655 \\
			\hline
			$0.025$  & 0.06175 -- 0.00057 i  & 0.06177 \\
			\hline
			\multicolumn{3}{||c||}{$Q=0.999Q_\text{max}$, $qQ=0.5$} \\
			\hline
			$\Lambda M^2$ & $\omega_\text{QNM}$ & $\omega_\text{peak}$  \\ [0.5ex] 
			\hline
			$0.005$ & 0.02599 + 0.00169 i & 0.02626 \\
			\hline
			$0.025$  & 0.06217 + 0.00004 i  & 0.06217 \\
			\hline
			\multicolumn{3}{||c||}{$\Lambda M^2=0.025$, $qQ=0.5$} \\
			\hline
			$Q/Q_\text{max}$ & $\omega_\text{QNM}$ & $\omega_\text{peak}$  \\ [0.5ex] 
			\hline
			$0.2$ & 0.06173 -- 0.00078 i & 0.06176 \\
			\hline
			$0.7$  & 0.06182 -- 0.00034 i  & 0.06183 \\
			\hline
			\multicolumn{3}{||c||}{$\Lambda M^2=0.025$, $Q=0.5Q_\text{max}$} \\
			\hline
			$qQ$ & $\omega_\text{QNM}$ & $\omega_\text{peak}$  \\ [0.5ex] 
			\hline
			$0.8$  & 0.09876 -- 0.00518 i & 0.09947 \\
			\hline
			$1$  & 0.12174 -- 0.00925 i & 0.12382 \\
			\hline
		\end{tabular}
	}
	\caption{Superradiant $l=0$ massless charged scalar QNMs $\omega_\text{QNM}$ in RNdS spacetime with various parameters and the respective frequency peak position $\omega_\text{peak}$ of the amplification factor $Z_0$.}
	\label{table}
\end{table}

First, we focus on $\ell=0$ waves which generally lead to superradiantly stable and unstable QNMs \cite{Zhu:2014sya,Konoplya:2014lha,Destounis:2019hca}. In Figs. \ref{varL}, \ref{varqQ}, \ref{largeqQ} and \ref{varQ} we show the amplification factors for $\ell=0$ waves, in various scenarios of the available parameter space. By fitting the amplification factor curves we find when $Z_0=0$ in order to further validate the accuracy of our numerical analysis. We find that the onset and termination of superradiance agrees very well with Eq. \eqref{suprad} for every case we explored.

\begin{figure*}[t]
	\includegraphics[scale=0.43]{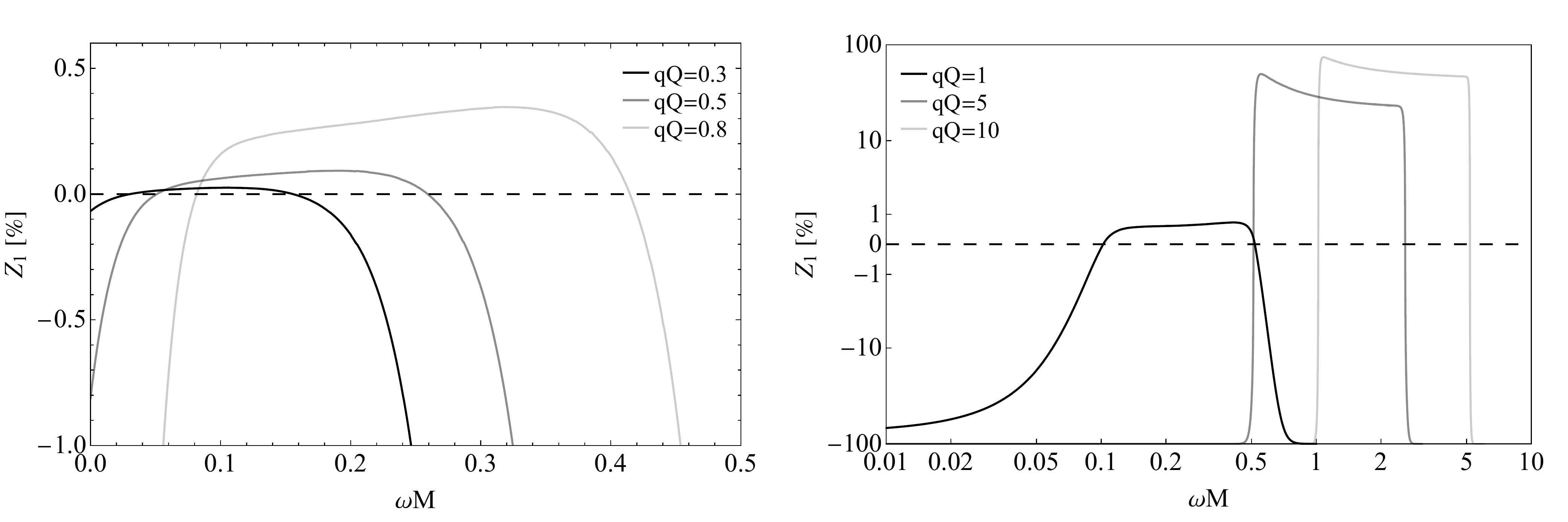}
	\caption{Left: Amplification factors of $\ell=1$ monochromatic waves for a RNdS BH with $Q=0.5Q_\text{max}$, $\Lambda M^2=0.025$ and varying charge coupling $qQ$. The horizontal dashed black line designates the onset of superradiant amplification. Right: Same as left but with larger charge couplings $qQ$.}
	\label{varqQl1}
\end{figure*}

\begin{figure}[t]
	\includegraphics[scale=0.42]{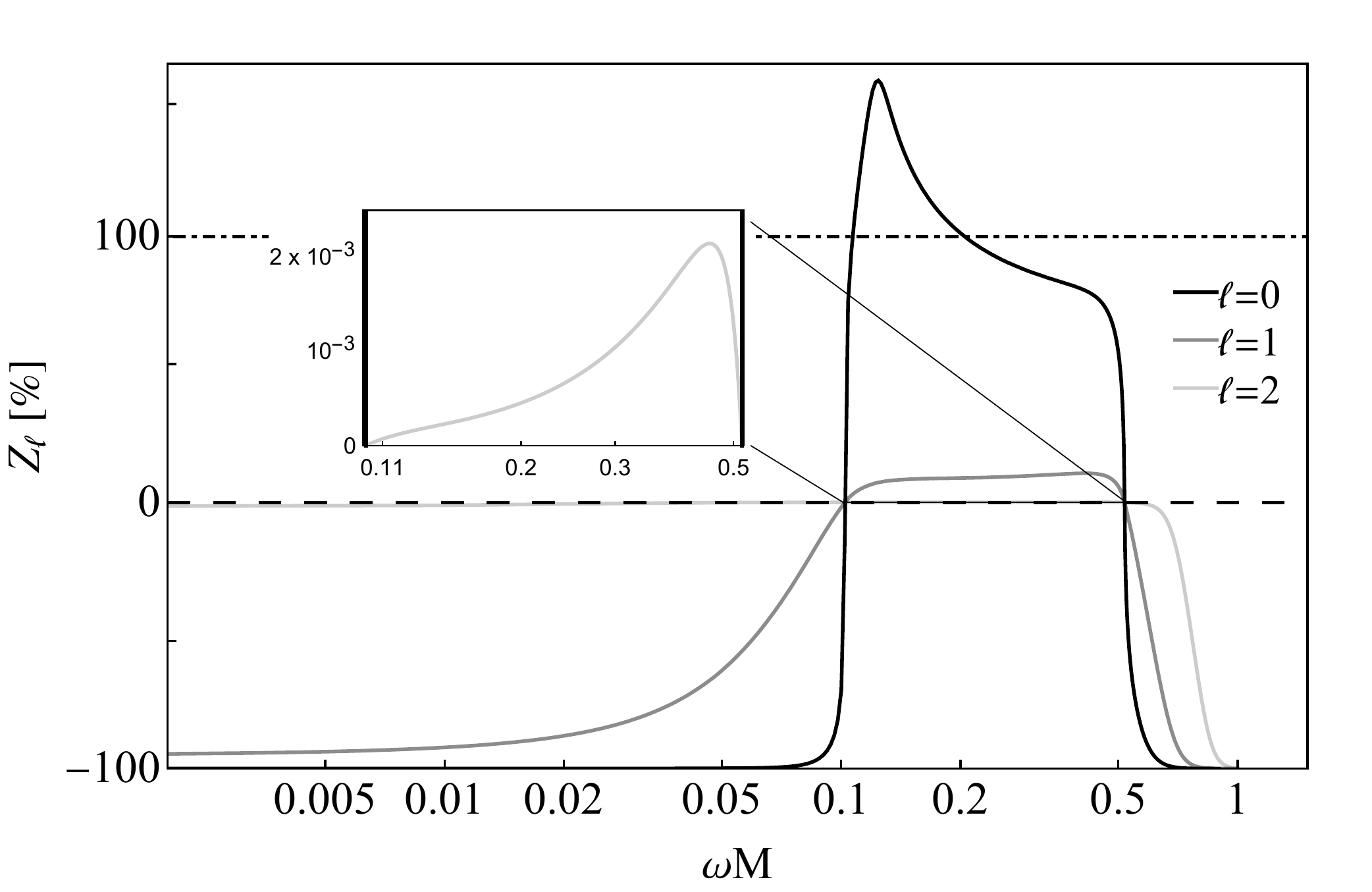}
	\caption{Amplification factors of monochromatic waves with charge coupling $qQ=1$ for a RNdS BH with $\Lambda M^2=0.025$, $Q=0.5Q_\text{max}$ and varying angular index $\ell$. The horizontal black dashed line designates the onset of superradiant amplification while the black dot-dashed line designates amplification factors that equal $100\%$.}
	\label{varl}
\end{figure}

In Fig. \ref{varL} we observe that even though the frequency domain of superradiance is shortened when a positive cosmological constant is present, in agreement with the findings in Kerr-de Sitter BHs \cite{Maeda:1993}, the amplification is significantly elevated beyond $100\%$ due to the existence of hyperradiating resonances. Here we have to note that the cosmological constants chosen to demonstrate our results are far from its physical values. This is due to the limitations of a numerical evolution with such a small value of $\Lambda$. Nevertheless, we can safely expect that the amplification factors $Z_{0}$ of RNdS BHs with arbitrarily small $\Lambda$ will be identical to that of RN BHs (see Fig. \ref{varL}). For any other sufficiently small but non-zero positive value of $\Lambda$, a resonant peak should be present in the superradiant regime with a height that
depends on the value of $qQ$ (the highest peaks will be present around the zeroes of imaginary part of the superradiant QNM in accord with the Breit-Wigner formula discussed below). The important aspect of arbitrarily small cosmological constants is the fact that the resonant frequency (real part of superradiant QNM) is placed arbitrarily close to $\omega M=0$ thus resonances should be visible (depending on the choice of $qQ$) at the infrared radiation regime.

The increment of the charge coupling $qQ$ shifts the superradiant regimes in larger frequency onsets as seen in Figs. \ref{varqQ}, \ref{largeqQ}. Our analysis demonstrates that at the large charge coupling limit $qQ>>1$, the amplification factors $Z_0$ approach $100\%$ from above, with ultraviolet monochromatic wave resonant peaks, in contrast to $Z_0$ in RN BHs which tends to $100\%$ from below \cite{Brito:2015oca}. Preexisting QNM studies (see e.g. Refs. \cite{Cardoso:2018nvb,Destounis:2019hca}) have shown that the increment of $qQ$ increases the oscillation frequency of QNMs monotonously, while when $qQ>>1$ then the absolute value of the imaginary part also increases. We can therefore conclude that at the large coupling limit the behavior of superradiance may be falsely assumed to be similar to that of RN BHs if one is unaware of the results presented on the whole parametric space involved in our analysis. In turn, the increment of the BH's charge enlarges the superradiant frequency regime and the amplification factors (see Fig. \ref{varQ}).

A careful fitting on the amplification curves, similar to that initially performed in \cite{Thorne1969}, unveils the underlying connection between the peak position, and its sharpness, and the $\ell=0$ scalar QNM resonance. The peaks appearing in the superradiant regime correspond to frequencies which lie close to the real part of the corresponding fundamental QNM of the system \footnote{Similar behavior has been observed in \cite{Khodadi:2020cht} where resonant peaks appear in the superradiant domain of scattered waves off of Kerr-Newman BHs in $f(R)$ modified theories of gravity.}. The extent of amplification and sharpness of the peak is analogous to how close (in absolute value) the imaginary part of the corresponding QNM is to the real axis. Therefore, even when the spacetime is stable under $\ell=0$ charged scalar perturbations, which generically occurs for sufficiently large cosmological constants and charge couplings \cite{Destounis:2019hca}, the superradiant amplification curve becomes sharper close to long-lived resonances and can lead to factors that are orders of magnitude larger than $100\%$, in accord with \cite{Thorne1969,ChandraFerrari1991,ChandraFerrari2}. Numerical tests have revealed that when the monochromatic wave's frequency hits a `sweet spot' for which the underlying spacetimes admits QNMs with extremely large lifetimes (which are practically normal modes up to numerical accuracy for specific charge couplings $qQ$) \cite{Cardoso:2018nvb,Destounis:2019hca}, the resonant hyperradiation peaks become delta functions (see also \cite{Kokkotas1994}).

The above picture is akin to Breit-Wigner resonances for which the incident waves that trigger trapped modes localized in the potential well behave, close to QNM resonances, as $|\mathcal{I}|^2\sim C\left[ (\omega-\omega_R)^2+\omega_I^2\right]$ \cite{Berti:2009wx}, where $C$ is a constant. Thus, when the frequency of the incident monochromatic wave matches that of the corresponding QNM resonance, the amplification factor is analogous to the lifetime of the resonance $\omega_I^{-2}$ (see Eq. \ref{amplification_factor}). 

In Figs. \ref{varL}, \ref{varqQ}, \ref{largeqQ} and \ref{varQ}, all cases depicted (besides the ones with vanishing $\Lambda M^2$) possess superradiant QNMs in their respective frequency domain. For completeness, we demonstrate the aforementioned resonant behavior in Table \ref{table}, where the peak position and the respective fundamental QNMs are shown for selected cases. 

\subsection{Higher multipole scalar waves} 

Higher angular index perturbations are not known to induce superradiant instabilities \cite{Zhu:2014sya}, neither support QNM resonances in the superradiant frequency regime. In this section we investigate if superradiance exists at all for $\ell>0$ in RNdS spacetime. As shown in Figs. \ref{varqQl1} and \ref{varl}, superradiant amplification of charged massless scalar waves can indeed occur for particular frequency ranges that strongly depend on the charge coupling. Our numerical analysis demonstrates that at the large coupling limit, where $qQ>>1$, the superradiant amplification tends to $100\%$ from below as in RN BHs. 

We further demonstrate how suppressed superradiance is for $\ell>0$ incident waves with the quadrupole ones reaching a menial amplification of $Z_2\sim 10^{-3}$ with respect to $Z_0,\, Z_1$ for a particular set of parameters (see zoomed region in Fig. \ref{varl}).

\subsection{Massive $\ell=0$ scalar waves} 

The role of the mass of scalar perturbations in RNdS is quite clear through the effective potential \eqref{RNdS_general potential}, as well as its effect on QNMs \cite{Cardoso:2018nvb,Destounis:2019hca}. The scalar mass antagonizes the charge coupling, which physically translates to a competition between the gravitational interaction of the BH mass and the massive field $\mu M$ and the electromagnetic repulsion of the BH's electric source and the charge of the field $qQ$ \cite{Konoplya:2014lha,Destounis:2019hca}. When the gravitational coupling $\mu M$ is very mild, the electromagnetic repulsion overcomes gravity and the superradiant instability remains, though is significantly suppressed. On the other hand, when the coupling $\mu M$ is strong enough, superradiantly unstable modes are completely absent from the QNM spectrum and, for $\mu M>>0$, no superradiantly stable QNMs exist at all \cite{Cardoso:2018nvb,Destounis:2019hca}. Therefore, one should expect that for sufficiently large $\mu M$, the amplification factors will never exceed $100\%$ since there will be no QNM resonances in the superradiant regime of monopole monochromatic incident waves.

\begin{figure}[t]
	\includegraphics[scale=0.42]{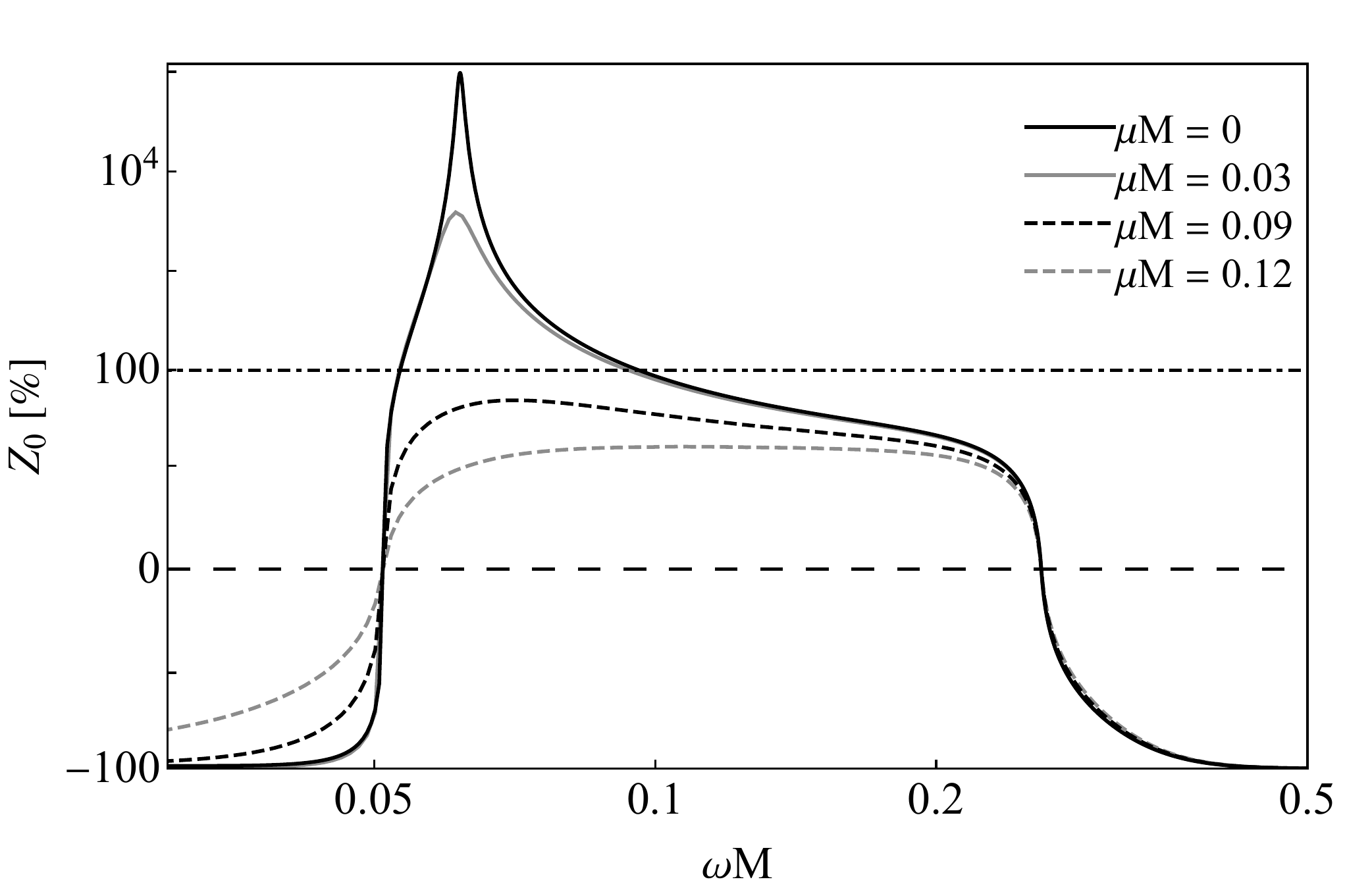}
	\caption{Amplification factors of $\ell=0$ massive monochromatic waves with charge coupling $qQ=0.5$ for a RNdS BH with $\Lambda M^2=0.025$, $Q=0.5Q_\text{max}$ and varying mass $\mu M$. The horizontal black dashed line designates the onset of superradiant amplification while the black dot-dashed line designates amplification factors that equal $100\%$.}
	\label{varmass}
\end{figure}

In Fig. \ref{varmass} we have depicted two cases for which the respective fundamental QNMs satisfy the superradiant relation ($\mu M=0, \,0.03$) and two more cases for which the corresponding fundamental QNMs are not superradiant ($\mu M=0.09, \,0.012$). It is noteworthy that the superradiant relation \eqref{suprad} does not depend on the mass of the field and that is clearly demonstrated in Fig. \ref{varmass} where the onset and termination of superradiance occurs exactly at the same frequencies regardless of the mass. It is clear that when there are no resonances in the superradiant domain, the amplification does not exceed $100\%$, but still occurs. Further numerical analysis unveils that even larger scalar masses diminish $Z_0$ quickly. 

\section{Conclusions}

In this work, we have explored the superradiant amplification of charged scalar incident waves on RNdS BHs. In generality, RNdS BHs superradiate in a similar manner as RN and Kerr BHs. Even so, new phenomena arise in RNdS due to the negative wells formed at the effective potential of charged scalar perturbations. Supperadiance in RNdS is severely amplified with respect to RN BHs, despite the fact that the frequency domain of superradiant amplification is lessened due to the introduction of a positive cosmological constant.

Ultralight and massless spherically-symmetric scalar waves can extract large amounts of electromagnetic energy from RNdS BHs (assuming an unrealistic cosmological constant) when the incident wave's frequency matches the real part of the corresponding fundamental QNM oscillation that satisfies the superradiant relation. The amplification factors in these cases are orders of magnitude larger than $100\%$ and loosely depend on the lifetime of the respective QNM which is dictated by its imaginary part, in accord with Breit-Wigner resonances \cite{Thorne1969,Kokkotas1994,Berti:2009wx}. Interestingly, the larger the lifetime of the superradiant QNM the higher the resonant amplification peak is regardless of the linear stability of spacetime. On the other hand, when the scalar field is massive enough, QNMs are expelled from the superradiant regime \cite{Destounis:2019hca} and superradiance is rapidly suppressed. 

Higher multipole incident waves do not exhibit resonant amplification since no QNMs reside in the superradiant regime. This leads to similar amplification factors as in RN, which asymptote to $100\%$ at the physical limit $qQ>>1$, in contrast to $\ell=0$ waves for which the amplification factors tend to $100\%$ from above at the same limit, even though ultraviolet resonances still occur but with small lifetimes. We can conjecture that the amplification of massive $\ell>0$ incident waves will also be highly quenched, therefore the most prominent amplifiers of scattered waves are linearly stable RNdS BHs with particular sets of parameters that lead to long-lived superradiant $\ell=0$ QNM resonances with arbitrarily large decay timescales. In any case, a physical choice of the cosmological constant should render the intriguing resonant hyperradiation phenomenon exhibited by RNdS BHs extremely ambitious, if not fictitious, in the majority of the parameter space involved expect when the charge coupling $qQ$ conspires to elicit extremely long-lived superradiant QNMs.

A very interesting extension of this work would be to perform an analysis of BH superradiance in charged accelerating spacetimes described by the $C$-metric \cite{Kinnersley:1970zw,Hawking:1997ia,Griffiths:2006tk}. Such spacetimes have a very similar causal structure with RNdS BHs, with the cosmological horizon being replaced by an acceleration horizon beyond which events are causally disconnected with the respective static region. Even though the $C$-metric is not spherically-symmetric, the neutral massless QNMs have been recently calculated \cite{Destounis:2020pjk,Destounis:2020yav}. Hence, the generalization to a charged QNM analysis, as well as the exploration of scattered charged scalar waves off charged accelerating BHs is of particular interest. We plan to pursue this work in the near future \cite{Destounis:2022rpk}.

\begin{acknowledgments}
The authors would like to warmly thank Vitor Cardoso for helpful discussions.
\end{acknowledgments}

\bibliography{references}

\end{document}